\documentclass[12pt]{article}
\usepackage{pic02}
\usepackage{url}
\usepackage{hyperref}
\usepackage{graphicx}

\newcommand{\etal}{\mbox{\textit{et al.}}}

\begin{document}

\title{\bf LARGE BULK MATTER SEARCH FOR FRACTIONAL CHARGE PARTICLES}

\author{
Irwin Lee \\
{\em Stanford Linear Accelerator Center}
}

\maketitle

\baselineskip=14.5pt

\begin{abstract}
We have carried out the largest search for stable particles with
fractional electric charge, based on an oil drop method that
incorporates a horizontal electric field and upward air flow.
No evidence for such particles was found, giving a 95\%~C.L.
upper limit of $1.15\times 10^{-22}$ particles per nucleon on
the abundance of fractional charge particles in silicone oil
for $0.18 e \le |Q_{residual}| \le 0.82 e$.
\end{abstract}

\baselineskip=17pt

\section{Introduction}

\begin{figure}[htbp]
  \centerline{\hbox{ \hspace{0.2cm}
    \includegraphics[width=6.5cm]{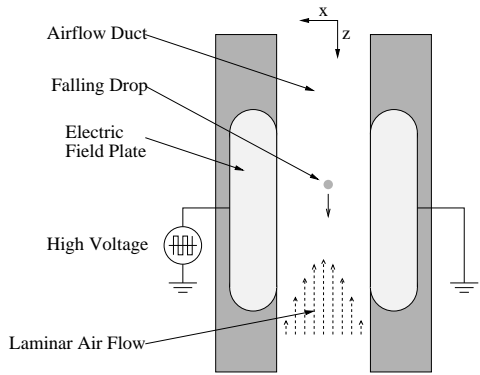}
    }
  }
  \caption{\it
	The experimental principle behind the search technique.
  \label{fig:principle}}
\end{figure}

We have carried out the largest search for fractional electric charge
elementary particles in bulk matter using 70.1~mg of silicone oil.  That is,
we looked for stable particles whose charge $Q$ deviates from $Ne$ where $N$
is an integer, and $e$ is the electron charge.  No evidence for such
particles was found.  We used our new
version \cite{loomba} of the Millikan oil drop method
containing two innovations compared to the classical method that we used in
Halyo \etal \cite{halyo}.  One innovation is that the drop charge is
obtained by observing the drop motion in a \emph{horizontal}, alternating
electric field compared to the classical use of a vertical electric
field.  The other innovation is the use of an upward
flow of air to reduce the vertical terminal velocity of the drop,
which enabled us to use larger drops, about
20.6~$\mu$m in diameter compared to the 10~$\mu$m drops used in our previous
experiments.  Figure~\ref{fig:principle} depicts the principle of the
technique.

\section{Apparatus}

Recent advances in inkjet printing technology and computer imaging
are the key enabling technologies behind the experiment.  The apparatus
is built around a micromachined \cite{patent},
drop-on-demand fluid ejector, and
a digital charge-coupled device (CCD) camera.  The drop ejector produces
extremely uniform oil drops with radii constant to $\pm 0.2\%$.  This
ability to precisely control the drop radius, in addition to the large
statistics gained by measuring many drops, allows the experiment to
be self-calibrating, significantly reducing the possibility of an artifact.

The experiment is completely automated.  The measurement region,
approximately 2.3~mm horizontally by 3.0~mm vertically, is viewed by the
CCD camera producing a 10~Hz stream of images in 8~bit greyscale at
$736 \times 240$ pixel resolution.  This
data is acquired, stored, and analyzed in real-time using custom written
software.  This software analyzes the images, measuring the positions
of the drops to sub-pixel accuracy, and performs pattern recognition
to identify the trajectories of individual drops.  The charge $Q$ on each
drop is calculated from the best possible fit to the measured trajectory,
where the precision is primarily limited by Brownian motion.

\section{Results}

Data was collected for 30 weeks.  From this sample, approximately $11\%$
of the drops were rejected using predetermined selection criteria.  These
selection criteria eliminated incorrectly measured drops, and were
unbiased with regard to fractional charge.  A total of $17\times 10^6$
drops were accepted into the final data sample.

\begin{figure}[htbp]
  \centerline{\hbox{ \hspace{0.2cm}
    \includegraphics[width=6.5cm]{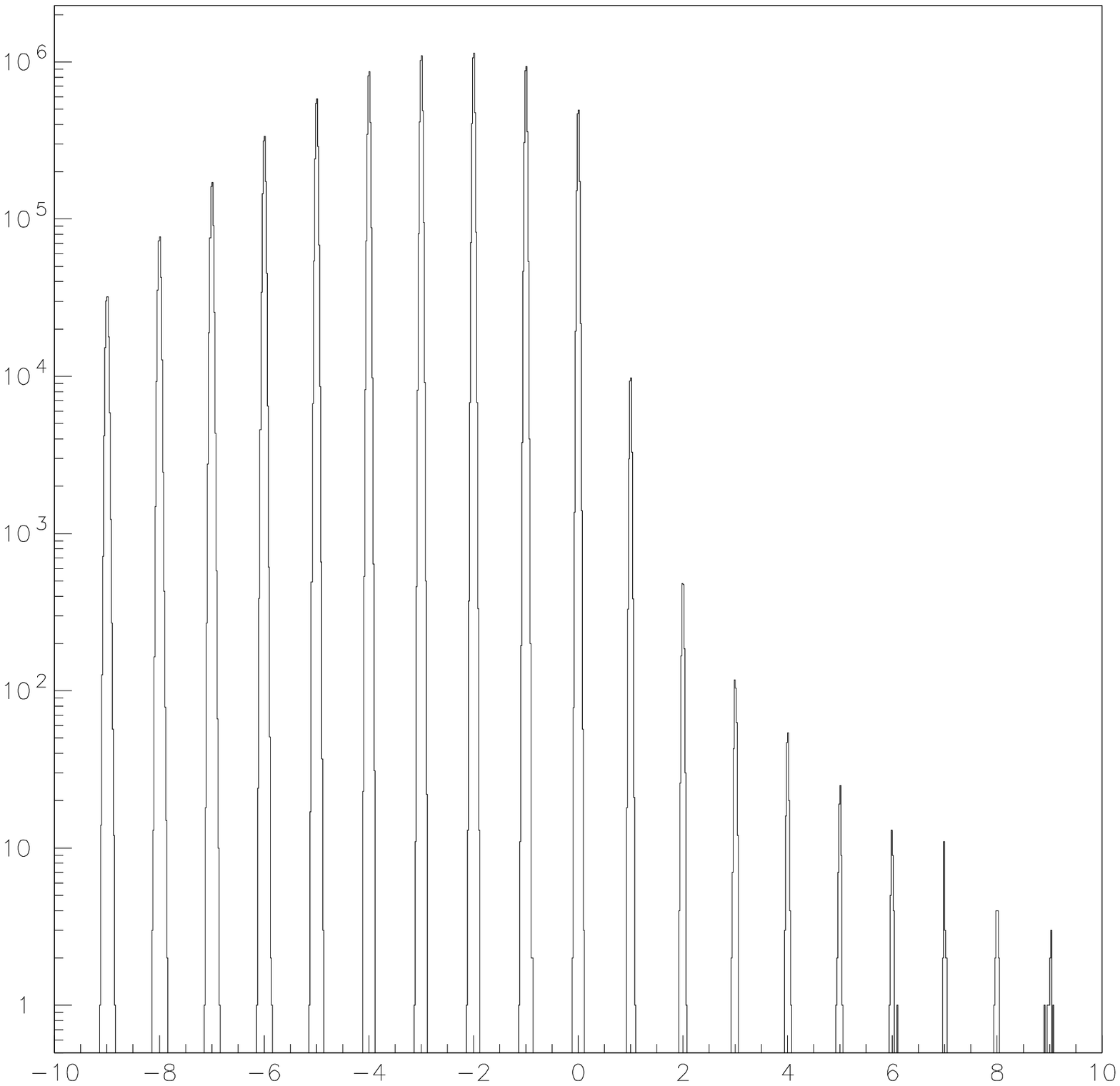}
    \hspace{0.3cm}
    \includegraphics[width=6.5cm]{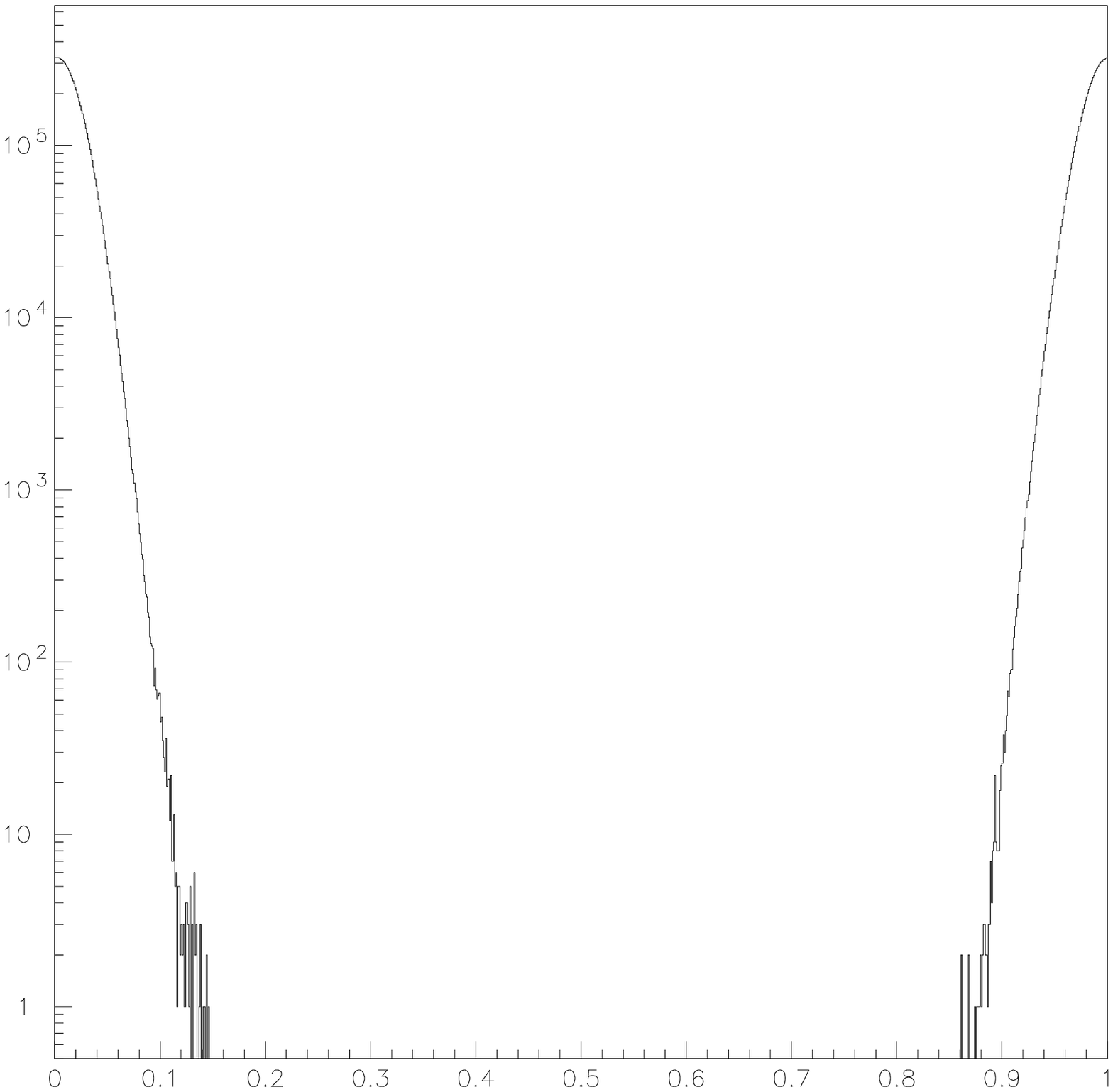}
    }
  }
  \caption{\it
	The overall charge distribution (left), in units of $e$, of the
$17\times10^6$ oil drops studied and the residual charge $Q_r$ (right).
  	\label{fig:results}
  }
\end{figure}

The resulting distributions of $Q$ and the residual charge $Q_r$,
defined to be $Q_r = Q-N_le$ where $N_l$ is the largest integer less
than $Q/e$, are shown in Figure~\ref{fig:results}.  No drops with
fractional charge were observed, yielding a 95\% confidence level
upper limit on the abundance of fractional charge particles in silicone oil,
for $0.18 e \le Q_r \le 0.82 e$, of $1.17\times 10^{-22}$ particles
per nucleon.  This result is described in detail in Lee \etal \cite{this}.

\section{Acknowledgments}

This experiment was performed at SLAC under
the direction of Martin Perl, and in collaboration with
Sewan Fan, Valerie Halyo, Peter C. Kim,
Eric R. Lee, Dinesh Loomba, and Howard Rogers.  Thanks also to Klaus
S. Lackner and Gordon Shaw.

This work was supported by Department of Energy contract DE-AC03-76SF00515.

\end{document}